\documentclass[twoside]{article}

\usepackage{amsmath,amssymb,amsfonts,amscd,epsfig}

\usepackage{color}
\usepackage{graphicx}

\newcommand{\be}{\begin{equation}}
\newcommand{\beq}{\begin{equation}}
\newcommand{\eeq}{\end{equation}}
\newcommand{\ee}{\end{equation}}
\newcommand{\bea}{\begin{eqnarray}}
\newcommand{\eea}{\end{eqnarray}}
\newcommand{\ba}{\begin{array}}
\newcommand{\ea}{\end{array}}

\newcommand{\ga}{\gamma}

\newcommand{\ep}{\epsilon}
\newcommand\IZ{\mathbb{Z}}
\newcommand\IR{\mathbb{R}}
\newcommand{\IC}{\mathbb{C}}
\newcommand\IT{\mathbb{T}}
\newcommand{\II}{\mathbb{I}}

\def \Sg {\Sigma}

\def \om {\omega}

\def \th {\theta}

\def \La {\Lambda}

\def \a {\alpha}

\def \ga {\gamma}

\def \sg {\sigma}

\def \ep {\epsilon}
\def \part {\partial}

\def \sqr#1#2{{\vcenter{\hrule height.#2pt
       \hbox{\vrule width.#2pt height#1pt \kern#1pt
          \vrule width.#2pt}
       \hrule height.#2pt}}}

\def\lsim{\mathrel{\rlap{\lower4pt\hbox{\hskip1pt$\sim$}}
    \raise1pt\hbox{$<$}}}         
\def\gsim{\mathrel{\rlap{\lower4pt\hbox{\hskip1pt$\sim$}}
    \raise1pt\hbox{$>$}}}         

\def\IR{{\hbox{{\rm I}\kern-.2em\hbox{\rm R}}}}
\def\IH{{\hbox{{\rm I}\kern-.2em\hbox{\rm H}}}}
\def\IC{{\ \hbox{{\rm I}\kern-.6em\hbox{\bf C}}}}
\def\IZ{{\hbox{{\rm Z}\kern-.4em\hbox{\rm Z}}}}
\def\rref#1{(\ref{#1})}
\begin{document}

\begin{center}
{\Large\bf
 2+1 dimensional gravity}\\
\vspace{.4in}
{J.\ E.~N{\sc elson}\\
       {\small\it Dipartimento di Fisica Sezione Teorica }\\
       {\small\it Universit\`a degli Studi di Torino, and}\\
       {\small\it I.N.F.N. Sezione di Torino,}\\
       {\small\it via Pietro Giuria 1, 10125 Torino, Italy}\\
 \vspace{.4in}      
{\sl A contribution to the forthcoming volume ``Tullio Regge:\\ an eclectic genius, from quantum~gravity to computer play'',\\ Eds. L~Castellani, A.~Ceresole, R.~D'Auria and 
P.~Fr\'e (World Scientific).}}
\end{center}
\vspace{.4in}
\noindent It gives me great pleasure to review some of the joint work by Tullio Regge and myself. 
We worked intensely on 2+1-dimensional gravity from 1989 for about five years, and published 16 articles. I will 
present and review two of our early articles, highlighting what I believe are the most important 
results, some of them really surprising, and discuss later developments.


\section*{\small Introduction} \label{intro} 

2+1-dimensional gravity is still a fascinating theory. One can study many issues of the four-dimensional theory without its technical complications, and the only other lower dimensional theory, two-dimensional gravity, has an action which is a topological invariant. It has a Chern-Simons formulation \cite{jac,achu,wit}, and the curvature is constant, proportional to the cosmological constant (positive, negative, or zero). There are no local degrees of freedom, only global topological ones, and only when there are non trivial topologies. However, it does have local exact solutions \cite{gar}, and admits three-dimensional black holes \cite{btz}. 

Often considered as a ``toy model'' without physical relevance, 2+1-dimensional gravity has, nevertheless, interesting connections to other theories, e.g. in the context of the Ads-CFT correspondence \cite{mald}, pure Chern-Simons theory on a manifold with boundary induces, for certain boundary conditions, the Wess-Zumino-Witten model \cite{wzw} on the boundary \cite{wit2}. For different boundary conditions, it induces Liouville theory at infinity \cite{henn}. Three-dimensional Chern-Simons theory can also arise as a string theory \cite{wit3}.

Tullio Regge always liked discrete physics, preferably with very few degrees of freedom, so when I showed him the 1989 Witten preprint on 2+1-dimensional gravity \cite{wit} which describes a theory with a finite number of only global topological degrees of freedom, he was very enthusiastic. We worked intensely on 2+1-dimensional gravity from then on, for about five years, and published 16 articles (see e.g \cite{NR7,NR1,NR2,NR8,NR0,NR5a,NR5b,NR3,NR4}). Here I review and complement two of our early articles \cite{NR7,NR1} which set the scene for many others. I will highlight some important results, e.g. the emergence of braid group relations, and the quantum group $SU(2)_q$, and discuss later developments. 

In both articles we studied the first order form of $3$ dimensional ($2$ space, $1$ time) de Sitter/anti de Sitter (AdS) gravity, where the gauge group is $SO(3,1)$ for cosmological constant $\Lambda >0$, and $SO(2,2)$ for $\Lambda<0$, and spacetime is $\IR \times \Sg_g$, $\Sg_g$ being a compact Riemann surface of genus $g$. In a previous article \cite{NR2} we had studied the Poincare case i.e. with $\Lambda = 0$,  where the gauge group is $ISO(2,1)$. We later showed \cite{NR8} that this can be obtained in the $\Lambda \rightarrow 0$ limit. 

In Section \ref{NR7} I discuss the first article \cite{NR7}, in which we constructed the algebra of observables for the torus $\IT^2$, $g=1$, and studied some of its representations. There are connections to knot theory, and to quantum groups. The quantum algebra of observables is the starting point for the second article \cite {NR1} which I discuss in Section \ref{NR1}. Here we studied  the symmetries of the quantum algebra of observables and how to generate them, implemented for any genus and at the quantum level. In Section \ref{later} I discuss further developments, in particular the algebraic structures which have been rediscovered in the mathematical and mathematical physics literature.

This text is intended as a companion to articles \cite{NR7,NR1}. It cannot be understood without having both of these to hand. 
Please note that in both articles the notation is unconventional. We used square brackets $[,]$ to denote Poisson brackets, and round brackets $(,)$ to denote commutators. I will refer to equations in both articles as equation (), but equations in this text simply as (). 

\section{\small Homotopy Groups and $2+1$ dimensional Quantum De Sitter
Gravity, J.~E.~Nelson, T.~Regge and F.~Zertuche, Nucl.\ Phys.\ B339 (1990) 516.}\label{NR7} 

In this article our construction used AdS connections $\omega^{ab}$ defined on page 517, and their Poisson brackets equation (1.4). These Poisson brackets and the AdS constraints $R^{ab}=d\omega^{ab}-\omega^{ac}\wedge\omega_c{}^b=0$ can be seen most clearly from the (2+1)-dimensional spacetime splitting $\IR \times \Sg$ of the Einstein action 
\beq
I_{\hbox{\scriptsize \it Ein}}
  = \int\!d^3x \sqrt{-{}^{\scriptscriptstyle(3)}\!g}\>
  ({}^{\scriptscriptstyle(3)}\!R - 2\Lambda) 
  = \frac{\a}{4} \int\!dt\int\!d^2x\,\epsilon^{ij}\epsilon_{abcd}\,
 (\omega^{cd}{}_j\,{\dot{\omega}}^{ab}{}_i-\omega^{ab}{}_0 R^{cd}{}_{ij}).
\label{bb1}
\eeq
In Section 2 we formally solved the constraints $R^{ab}=0$ (Einstein's equations) by the expression 
\beq
d\psi^{ab}=\om^{ac}\, \psi_c{}^b
\label{dpsi}
\eeq
where $\psi^{ab}$ is an $\hbox{SO}(3,1)$ or $\hbox{SO}(2,2)$-valued zero-form. We argued that \rref{dpsi}, when restricted to a path $\ga$ on $\Sg_g$, will lead to a set of $\psi^{ab}$ which depend only on the homotopy class of $\ga$ and its starting and end points. Joining these points (closing the path $\ga$) means that the new $\psi^{ab}$, now denoted $\psi^{ab}(\ga)$ depends only on the homotopy class of the closed path $\ga$, and the base point. i.e. these $\psi^{ab}(\ga)$ furnish a representation of the fundamental group $\pi_1$ of $\Sg_g$, whose presentation is given in equations (1.5) and (1.6). We also argued that, from equation (1.4), that the single elements of the $4 \times 4$ matrices $\psi(\ga)$ and $\psi(\sg)$, where $\ga,\sg ~\epsilon~\pi_1$ will have zero Poisson brackets iff they are obtained from integration of \rref{dpsi} over paths with zero intersection. We proved this in Section 3. This was the first hint of a quantum group structure though it differed from that of e.g.\cite{fadd} where elements of the same matrix may not commute but elements from different matrices do, the exact opposite of what we found. 

We calculated the Poisson brackets of the matrices $\psi(\ga)$ and $\psi(\sg)$ by expanding them along infinitesimal paths, see equation (3.2). The result is given in equations (3.4), and the r.h.s indeed depends on the intersection number $s$. It is valid only when $\rho, \sg$ have a single intersection i.e. $s=\pm 1$. Clearly the sign of $s$ depends on the orientation of the paths $\rho, \sg$.

The spinor version of Section 4 is more useful, where the spinor equivalent of $\psi(\ga)$ are $2 \times 2$ matrices $S^{\pm}$. They are related through equation (4.3). Their Poisson brackets are given in equations (4.8). Details can be found in Appendix A. The $\pm$ refer to the two copies of the $S^{\pm}$. For $SO(2,2)$ ($k=-1, \La <0$) the $S^{\pm}$ are real and independent, i.e. $S^{\pm}~ \ep~ SL(2,\IR)$, whereas for $SO(3,1)$ ($k=1, \La >0$) they are complex conjugates, i.e. $S^{\pm}~ \ep~ SL(2,\IC)$. 

Equations (4.8), though messy, are highly suggestive. In the first, the r.h.s depends on the intersection number $s$. Note also that $\rho_1,\sg_1$ denote the segments of the paths $\rho,\sg$ before the intersection, and $\rho_3,\sg_3$ denote the segments after (the paths are divided as $\rho=\rho_3\rho_2\rho_1$ and similarly $\sg$). The infinitesimal segments with a single intersection are $\rho_2,\sg_2$. 

Viewed in this way, if e.g $\rho$ is the path top left to bottom right and $\sg$ is the path bottom left to top right (with both start and endpoints identified), then the first of equations (4.8) can be represented graphically as the classical bracket between the paths $\rho,\sg$, shown in Figure 1.
\begin{figure}
\begin{center}
\includegraphics[width=20pc]{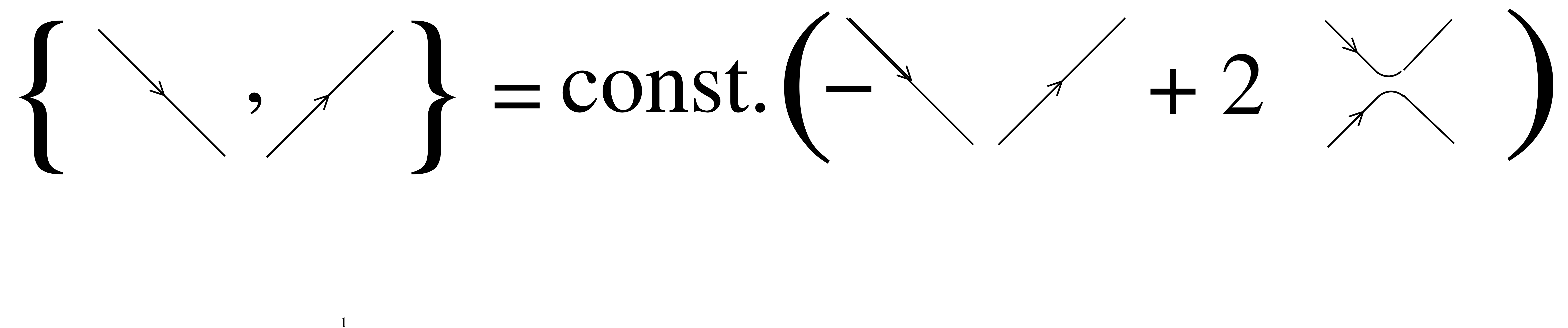}
\end{center}
\label{rog} 
\caption{The Poisson bracket of the paths $\rho, \sg$.}
\end{figure}

This was a surprise, but we had seen similar relationships in knot theory. In fact, we later realised that Figure 1 is highly reminiscent of the Kauffman polynomial \cite{kauff} for classical knots and links, and that it had already appeared in \cite{gol}. We extended the relationship to knot theory in \cite{NR0,NR5a} where we constructed, for genus $2$, a group of automorphisms of $\pi_1(\Sg)$ and a group of canonical transformations on the classical algebra of observables. Its symmetries are those of the braid group for $6$ (=$2g+2$) threads.

In Section 5 taking traces, equation (5.1), ensures gauge invariance, where a gauge transformation on a path $\rho$ is $\rho \rightarrow \nu \rho {\nu}^{-1}$ for any open path $\nu$, i.e. a shift of the base point. For 2+1-dimensional gravity this is equivalent to a diffeomorphism (coordinate transformation). The traces equation (5.1) are clearly gauge  invariant. The result is the classical non-linear algebra, equation (5.2). There is another version, equation (5.4), which Regge and I did not study. It is linear, but subject to non-linear constraints (equation (5.3)).\footnote{Roger Picken and I have studied this second version, and it has led us to a new quantum group structure, some interesting properties in quantum geometry, and a quantisation of the Goldman bracket \cite{np}.} In equation (5.2) $u,v$ refer to two oriented closed paths (cycles) with a single intersection on $\Sg_g$, and there are three independent traces $c(u),c(v),c(uv)$. Intersection $\pm 1$ is sufficient for the torus $g=1$, and the rest of \cite{NR7} discusses, for the $+$ algebra, the consequences of equation (5.2), its quantisation, and a preliminary study of its representations. 

The three traces in equation (5.2) are denoted, for the $+$ algebra, $x=c^+(u),y=c^+(v),z=c^+(uv)$, and their Poisson brackets are given in equation (6.1). They have a remarkable triangular, or cyclical structure, and were easy to quantise in Section 6. Firstly define the commutators as $i \hbar$ times the r.h.s. of the classical brackets equation (6.1) and replace the product $xy$ with $\frac{xy+yx}{2}$. The result is the commutator, equation (6.2), which I repeat
\beq
(x,y) = xy - yx = e^{i\th}~(z-xy) 
\label{comm1}
\eeq
and cyclical permutations of $x,y,z$. The parameter $e^{i \th}$ depends on $\Lambda$, $\hbar$, and the intersection number $s$. The $-$ algebra is identical to \rref{comm1} but with $\th \rightarrow -~\th$. Equation (6.2) is rearranged more symmetrically in equation (6.3). This was the second surprise. In fact equation (6.3) is referred to as the cyclical presentation of $SU(2)_q$, with $q = e^{i\th}$ \cite{fair,maj}. 

We also found some preliminary isometries of the algebra equation (6.3), e.g. equation (6.6) corresponds to the Dehn twist known as
$S:  U \rightarrow V^{-1}, V\rightarrow U$. We later extended this discussion to any genus in \cite{NR3}. 

We found an important central element, equation (6.4) (actually two, one for each $\pm$ algebra). In the classical limit $\th \rightarrow 0$ it can be expressed, using equation (5.3), as 
\beq
F^2= 1 +2xyz-x^2-y^2-z^2=tr (~\II S(U)~S(V)~S(U^{-1})~S(V^{-1})~)
\label{fclass}
\eeq
and therefore $F^2=0$ can be considered as an application of the group identity $UVU^{-1}V^{-1}=1$ to the representations $S$. It is worth noting that, considering the two $\pm$ algebras, there are actually six traces $x^{\pm}, y^{\pm},z^ {\pm}$, so the theory is overprescribed. But there are two central elements ${(F^2)}^{\pm}$, leaving four independent variables, which characterise the two degrees of freedom we wished to describe. Further, setting $F^2=constant$ \rref{fclass} defines a cubic surface, and we used that to classify the various solutions, showing that only the unbounded (hyperbolic) solutions correspond to an $SO(2,2)$,  $\Lambda<0$ theory, whereas the bounded (parabolic) solutions correspond to the unphysical $SO(4)$. In fact it was later shown in \cite{Ez2} that, by considering both $\pm$ algebras, of the nine possibilities (hyperbolic, elliptic, parabolic for each algebra), only the ``hyperbolic-hyperbolic'' solutions are physical. For solutions with elliptic or parabolic holonomies, spacetime still has the topology $\IR \times \IT^2$, but the toroidal slices are not spacelike.

In Sections 7 and 8 we discussed the bounded and unbounded representations for $\La < 0$, and those for $\La > 0$, respectively.

The unphysical bounded representations corresponding to $SO(4)$ are a direct generalisations of the theory of angular momentum. Using raising and lowering operators $K^{\pm}$, equation (7.1), and with $z  = sin \mu$, $\mu^{\dag} = \mu$, $\vert z \vert <1$ then $z, K^{\pm}$ satisfy the commutators of $SU(2)_q$.  The quantum Casimir $F^2$ , equation (6.4), is written in terms of these variables and finite-dimensional representations are constructed. There are a number of results, all incomplete, but one is worth noting, i.e. when $\frac{\pi}{\th} = q$ is an integer. Then, from the Weyl representations equation (7.15), the products $A^q,B^q,C^q$ of the unitary operators $A,B,C$ lie in the center of the algebra. This is typical of quantum algebras \cite{root}.

Unbounded representations are obtained by letting $\mu$ be non-hermitian. With $z^{\dag} = z$ it follows that $\mu^{\dag} + \mu = \pi$, and $\mu = \frac{\pi}{2} + i \nu$, $\nu$ real, with conjugate $\zeta$. The Weyl representation has hermitian operators $A, B, C$.

The de Sitter ($\La > 0$) algebra of Section 9 is similar, but with two sets of complex conjugate traces, and $\th$ replaced by $-i\th$. The limit $\th \rightarrow 0$ corresponds to the linear representations of $SO(3,1)$ \cite{nai}.

\section{\small $2+1$ Quantum Gravity, J.~E.~Nelson and T.~Regge, Phys. Lett. B272 (1991) 213.}\label{NR1}

In this short letter \cite{NR1} we streamlined many things, the presentation is pretty tight, and, as we said in the abstract ``Here all of the previous results are implemented for any genus and at the quantum level''. It can be considered as a prelude to \cite{NR4} where we studied the representations of the $g=2$ theory in some detail.

We extended the quantum algebra equation (6.3) of \cite{NR7} or \rref{comm1} to paths with more than one intersection, for $\La<0$ (and setting $s=1$). To this end we used a different representation of $\pi_{1}$ of $(\Sg_g)$ for genus $g$. It has generators $t_{i}$, $i=1\cdots 2g+2$, related to the $U_i,V_i, i=1...g$ in equation (6), and they satisfy the three identities in equation (7). The $t_i$ run around the perimeter of the polygon, Figure 1 of \cite{NR1}, but there are also diagonals (compound paths) like e.g. $d_{37}$, which is homotopic to $t_3t_4t_5t_6$. One of the advantages of using the the $t_i$ is that the number of intersections is at most $\pm 2$ \cite{NR5b}. Two paths, either perimeter or compound, that ``touch'' on the border of the polygon have intersection number $\pm 1$, those that do not touch anywhere have zero intersection, and those that cross inside the polygon have intersection $\pm 2$. The quantum traces  corresponding to the paths $d_{ij}$ are denoted $a_{ij}= c(d_{ij})$. For intersection $\pm 1$ the commutator \rref{comm1} applies, and for zero intersection the commutator is zero. Many of the commutators of traces corresponding to paths with intersection $\pm 2$ were obtained by using the programme Mathematica, the others by repeated use of the quantum Jacobi identity $(A,(B,C)) + (B,(C,A)) + (C,(A,B))=0$. These are reported in equations (12) and (13), where $K= e^{i\th}$.

The quantum algebra equation (6.3) is highly overprescribed, e.g. for $g = 2$ there are 15 independent $a_{ij}$, but in \cite{NR3}, we showed that for the classical algebra, equation (5.2) of \cite{NR7}, by using a set of rank and trace identities, that there are precisely $6g - 6$ independent traces, the number of independent moduli on a Riemann surface of genus $g$.

We also found a set of generators for the quantum mapping class group. In analogy with the quantisation for $g=1$ in \cite{NR7} we parametrised one trace, $a_{jk}$, for $j,k$ fixed by $a_{jk} = \frac{\cos \psi}{\cos \frac{\th}{2}}$ (the analogue of $z = sin \mu$) where, for hyperbolic solutions, $\psi$ is not hermitian, and therefore functions of $\psi$ are not periodic. We looked for functions of $\psi$ that generate, through their commutators, transformations on all the other $a_{ij}$ that have non-zero commutators with this $a_{jk}$, by starting with a triangular structure (the analogue of $x,y,z$). For the triangular structure the intersections are $\pm 1$ but were easily extended to intersections $\pm 2$. The result is that the quantum counterpart of the classical Dehn twists is generated by conjugation with Gaussians of $\psi$, see equations (17-19), by using the BCH formula
\beq
e^{A} B e^{-A}= B + [A,B] + \frac{[A,[A,B]]}{2!} + \dots
\label{bch}
\eeq


\section{\small Later Developments}\label{later}

In a later article \cite{NR4} we studied the representations of the $g=2$ theory in some detail. We were able to reduce the $15$ independent traces to $6$ ($6g-6$) by using choosing three commuting traces, parametrised by three angles, and their conjugate momenta. This is highly non-unique, but we were also able to obtain formulae for the remaining $12$ traces. 

Waelbrook and Zertuche have considered extensions to conformal gravity, supergravity, and introduced a time variable (one of the traces) \cite{zert}.

With other collaborators I have studied the relationship between this approach and the second order metric formalism of Moncrief and others \cite{vm}, and the linear version of the algebra, equation (5.4) of \cite{NR7}.

Steve Carlip and I compared the second order ADM metric formalism to the first order holonomy formalism for the torus, showing that they can be considered as ``Schr{\"o}dinger'' and ``Heisenberg'' pictures, respectively.  With a hyperbolic representation of the algebra \rref{comm1} the holonomy parameters are related to the torus modulus and momentum through a time-dependent canonical transformation. The quantisation of the two classically equivalent formulations differs by terms of order $O(\hbar^3)$, negligible for small $\vert \La \vert$. Further, the quantum modular group splits the holonomy representation Hilbert space into physically equivalent orthogonal ``fundamental regions'' that are interchanged by modular transformations \cite{cn}.

Vincent Moncrief and I used linear combinations of constants of motion to satisfy the AdS algebra $\hbox{so}(2,2)$ 
in either ADM or holonomy variables. Quantisation is straightforward in terms of the holonomy parameters, and the 
modular group is generated by these conserved quantities. On inclusion of the (time-independent part of) the Hamiltonian three new global constants are derived, and the quantum algebra extends to the conformal algebra $\hbox{so}(2,3)$ \cite{nm}. 

Roger Picken and I have studied the linear algebra equation (5.4) of \cite{NR7}, which Regge and I did not study. It is linear, but subject to non-linear constraints (equation (5.3)). It has led us to a new quantum group structure, some interesting properties in quantum geometry, a quantisation of the Goldman bracket \cite{gol}, and a theory of intersecting loops on surfaces \cite{np}.

The algebraic structures that we found, both classical and quantum, have been rediscovered and developed in the mathematical and mathematical physics literature, in particular their symmetries and invariants. I give just a few examples. 

For $g=1$ the classical cubic Casimir (4) is the Markov polynomial \cite{mol}, and the elements $x,y,z$ have been used as elements of $3 \times 3$ Stokes matrices \cite{dub}. 

For $g=2$ we constructed an explicit representation of the mapping class group acting on $\pi_1(\Sg)$ \cite{NR0,NR5a}. It was reviewed in \cite {NRzent}.

For arbitrary $g$ our set of mapping class group induced observables corresponds to the centraliser of a non-trivial mapping class group \cite{baad}. The quantum algebra, equations (13) of \cite{NR1} is isomorphic to the non-standard $q$-deformed algebra ${U_q}^{\prime}(so_n)$, $n=2g+2$  \cite{gav}, and can be obtained by quantising the coordinates of Teichmuller space \cite{chf}. Further, the quantum algebra has been the subject of 2 UK EPSRC Research grants \cite{mazz}. It also appears in Tropical Geometry \cite{chm}.

\section*{\small Final Remarks}

In this first-order approach, the constraints have been solved exactly.  There is no Hamiltonian, and no time development. One can think of this formalism as a ``Heisenberg picture'' time-independent description of the entire spacetime, or as initial data for some choice of time. It is related, for the torus, to the time dependent ADM theory \cite{cn}. 

We only studied the $+$ algebra but a physical theory would need both $\pm$ copies of the quantum algebra equation \rref{comm1}.  Indeed, when Carlip and I made the connection to the ADM theory, for the torus \cite{cn}, we used a hyperbolic representation of the three traces $x, y, z$ which satisfy (3), with parameters $r_1^{\pm},_2^{\pm}$. The unique action of (2+1)-dimensional gravity can then be written as $\int \a(r_1^-dr_2^- - r_1^+ dr_2^+)$.

Tullio Regge had clearly been thinking about the relationship between discrete geometry, topology, and higher dimensions for a long time. He often talked about his dream of a ``swiss cheese universe'' \cite{tr}, the torus, or doughnut ($g=1$) and a pretzel ($g=2$) being low genus examples. His idea was that a lower dimensional system with very high genus $g$ could mimic higher dimensions. For example, a discrete scaffolding of a three-dimensional space with ``fattened pipes'' can be considered as an approximation to a high genus two-dimensional surface $\Sg_g$. On $\Sg_g$ there would be a set of closed loops (cycles), each cycle circling either a hole or a pipe. In this scenario the curvature would be constant on $\Sg_g$, and the cycles should satisfy certain identities, e.g. equation (1.6) of \cite{NR7} or equation (7) of \cite{NR1}. 

This idea was considered in the context of loop quantum gravity \cite{rovsmol}. There is a noncanonical graded Poisson algebra of classical observables, which are nonlocal and relate to loops in a three-manifold $\Sg$. The quantum theory involves a deformation of this algebra.

More recently a discretised and complex Chern-Simons theory was quantised, and led to a four-dimensional model with non-zero $\La$ \cite{hagg}. 

Leonid Chekhov, Regge and I made a preliminary attempt to study the large genus limit. As a first step, we expressed the observables in terms of Penner-Fock coordinates, and used a graph description with a discrete, finite number of components. By considering continous non-discrete components in a scaling limit some elements of the algebra become commuting, the remaining 2 continuous elements satisfying Poisson brackets for string coordinates. All of the observables can be expressed in terms of these \cite{cnr}.

It was a great pleasure and privilege to have known, and worked with Tullio Regge, the most brilliant scientist I ever met. We shared moments of enlightment, and the pure joy of discovery. Thank you, Tullio.

\section*{\small Acknowledgements}

\noindent This work was supported by INFN - Istituto Nazionale di Fisica Nucleare - Iniziative 
Specifiche GSS, and MIUR - Ministero dell'Istruzione, Universit\`a e della Ricerca Scientifica e 
Tecnologica. I am grateful to the University of Turin for one year of teaching leave, and to Roger 
Picken for his help in preparing the figure for the braid group relation Figure \ref{rog}.

\bibliographystyle{}

\end{document}